\documentclass[preprint,12pt]{elsarticle}

\usepackage{amssymb}
\usepackage[table,xcdraw]{xcolor}
\usepackage{amsmath}
\usepackage{booktabs}
\usepackage{siunitx}
\usepackage{physics}
\usepackage[draft]{changes}
\usepackage{cleveref}
\usepackage[version=4]{mhchem}

\usepackage{lineno}

\journal{}

\AtBeginDocument{\RenewCommandCopy\qty\SI}
\begin{document}

\begin{frontmatter}



\title{Coupled Effects of Biofilm Dynamics and Flow on Hydrogen Transport in a Hydrogen-Generating Fracture} 

\author[label1]{Siqin Yu}
\author[label2]{Yinuo Noah Yao\corref{cor1}} 
\ead{yyao@tamu.edu}
\cortext[cor1]{Corresponding author}
\affiliation[label1]{organization={Department of Civil and Environmental Engineering},
            addressline={University of California, Los Angeles}, 
            city={Los Angeles},
            state={CA},
            postcode={90095},
            country={USA}}
\affiliation[label2]{organization={Zachry Department of Civil and Environmental Engineering},
            addressline={Texas A\&M University}, 
            city={College Station},
            state={TX},
            postcode={77843}, 
            country={USA}}

\begin{abstract}
Biofilm growth in hydrogen-generating fractures modifies fluid flow and hydrogen transport, while microbial activity consumes hydrogen. These coupled processes influence hydrogen recovery in stimulated geologic hydrogen systems. We use high-fidelity simulations to investigate flow, biofilm dynamics, and hydrogen transport in a rough fracture with reactive walls. Flow rate and biofilm permeability were varied to quantify changes in fracture permeability, biomass accumulation, and outlet hydrogen concentration. Biofilm growth reduced apparent fracture permeability by up to 65\% as biomass occupied the aperture. Two hydrogen concentration regimes are identified. For highly permeable biofilms (Type I), outlet concentration evolved non-monotonically as roughness-induced vortices initially restricted hydrogen transport before biofilm growth suppressed them. Low-permeability biofilms (Type II) result in an approximately monotonic outlet concentration decrease dominated by transport resistance and microbial consumption. The regimes were distinguished by a Darcy-scale Reynolds number ($Re_D=u\sqrt{K_0}/\nu$). Temporal hydrogen concentration variations in field studies may help determine intrinsic biofilm permeability.
\end{abstract}


\end{frontmatter}
\section{Introduction}

Hydrogen has attracted increasing interest as a clean energy fuel and carrier \cite{zivar2021underground}. In subsurface environments, hydrogen can be generated through water-rock reactions (e.g., serpentinization of olivine-bearing minerals) \cite{mao2025geologic,templeton2024low,yuan2026stimulated}. Hydrogen also serves as an electron donor for microbial processes, including sulfate reduction \cite{han2016mineral,liu2025growth}, methanogenesis \cite{hellerschmied2024hydrogen,chapelle2002hydrogen}, and homoacetogenesis \cite{thaysen2021estimating,molenaar2017competition}. These microbial reactions consume hydrogen and alter local aqueous chemistry, while microbial growth, particularly biofilm formation, can modify flow and mass transport within the pore space.

In stimulated geologic hydrogen systems, injected fluids react with fresh mineral surfaces exposed along fractures, generating hydrogen through water-rock reactions. During serpentinization, ferrous iron ($Fe^{2+}$) is oxidized to ferric iron ($Fe^{3+}$) while water is reduced to hydrogen (H$_2$). Low-temperature stimulated hydrogen systems are particularly attractive because they may be operated at relatively shallow depths ($<3$ km), potentially reducing drilling and stimulation costs \cite{templeton2024low,corre2023quaternary,lukawski2014cost}. At these depths, microbial communities may remain abundant and metabolically active, particularly where hydrogen is available \cite{boden2025evaluating,momper2017energy,bell2022active}. The recoverable hydrogen amount may therefore depend not only on water-rock reaction rates but also on microbial consumption and the effects of biofilm growth on fluid flow and mass transport \cite{ellis2024model}.

In hydrogen-generating fractures, microbial activity can affect hydrogen recovery through two coupled mechanisms: direct hydrogen consumption and biofilm-induced changes in flow and transport. Microbial hydrogen consumption has been recognized as an important process in underground hydrogen storage and other hydrogen-rich subsurface systems \cite{liu2023pore,dopffel2023microbial,thaysen2021estimating}. Experimental and modeling studies have shown that hydrogen consumption depends strongly on local transport conditions, which control the supply of H$_2$ and nutrients to active microorganisms \cite{rolland2025impact,strobel2023coupled,mahmoodi2026hydrogen,yu2024analytical}. In turn, hydrogen availability and consumption can influence microbial growth and population dynamics \cite{liu2025growth,bagnoud2016reconstructing,mura2025microbiological}.

In addition to existing as freely suspended cells, microorganisms in subsurface environments can grow as biofilms attached to mineral grain or fracture surfaces \cite{phillips2013potential,kurz2022competition}. Biofilms consist of microbial communities embedded in a matrix of extracellular polymeric substances, forming a porous layer between the hydrogen-generating surface and the bulk fluid. Within this layer, hydrogen can be transported by diffusion and advection while being consumed by metabolically active microorganisms \cite{zhou2022rates,kurz2023morphogenesis}. As biofilms grow, they can also alter the fracture geometry by reducing the flow aperture, modifying local velocity fields, and redistributing shear stresses \cite{kurz2022competition,deng2013effect,markale2023internal,Shi2026-zc}. 

The extent to which biofilms affect hydrogen transport depends on both their thickness and intrinsic permeability. While biofilm thickness determines the distance over which hydrogen must travel from the reactive wall to the bulk fluid, its permeability controls the extent of advective transport within the biofilm. For a given thickness, a low-permeability biofilm suppresses advective exchange with the bulk fluid, making hydrogen transport increasingly dependent on molecular diffusion. In contrast, a highly permeable biofilm allows greater fluid penetration, making hydrogen transport more sensitive to the local flow field. Previous studies have shown that biofilm permeability affects the flow field, shear-stress distribution, and solute transport in bioclogged porous media \cite{deng2013effect,peszynska2016biofilm}. In particular, Deng et al. showed that biofilm permeability influences the distribution of shear stresses and the shape of solute breakthrough curves \cite{deng2013effect}. More recently, Dawi et al. demonstrated that the relationship between porosity and effective permeability is strongly influenced by the prescribed biofilm permeability \cite{dawi2024pore}. Kurz \emph{et al.} \cite{kurz2023morphogenesis} performed microfluidic experiments to investigate the evolution of biofilm morphology and associated hydrodynamics, revealing spatial variations in the estimated permeability of the biofilms. Despite its importance, local biofilm permeability remains difficult to measure directly under in situ conditions and is often prescribed or calibrated in numerical models.

The coupling between biofilm growth and hydrogen transport is particularly relevant to stimulated geologic hydrogen systems, where injected fluids can access fresh iron-bearing mineral surfaces through the stimulated fracture network \cite{yuan2026stimulated}. Flow through rough fractures can exhibit complex velocity distributions, including local recirculation and eddies at sufficiently high Reynolds numbers \cite{briggs2017numerical,wang2020complexity}. These flow structures may influence the transport of hydrogen generated at the fracture walls and its exchange with the bulk fluid. When biofilms grow on the same reactive surfaces, hydrogen must be transported through or around the biofilm before it can be recovered, during which a fraction of hydrogen may be consumed by the microorganisms. Biofilm growth can, in turn, modify the local flow field and the transport of hydrogen away from the reactive surface. Numerical methods have also been developed
to simulate reactive transport in complex geometries
with moving interfaces \cite{Yousefzadeh2023-rg}. Although the effects of biofilm growth on flow and solute transport have been investigated extensively, most previous studies have considered reactants supplied by the flowing fluid rather than generated at the solid surface. It therefore remains unclear how fracture flow and biofilm permeability jointly control microbial growth, hydrogen consumption, and the produced hydrogen concentration in hydrogen-generating fractures. 

In this study, we investigate the coupled dynamics of fracture flow, biofilm growth, and hydrogen transport in a hydrogen-generating fracture. We develop a high-fidelity numerical model that resolves fluid flow in both the open fracture and the porous biofilm, hydrogen generation at the fracture walls, microbial consumption, and the evolution of biofilm biomass over approximately 12 days. By systematically varying the Reynolds number in the fracture and intrinsic biofilm permeability, we quantify their effects on apparent fracture permeability, total biofilm biomass, active biofilm fraction, and outlet hydrogen concentration. We identify two distinct hydrogen-transport regimes associated with high-permeability and low-permeability biofilms and characterize the transition between these regimes.

The remainder of this paper is organized as follows. Section~\ref{method} presents the problem formulation (Section~\ref{sec:problem_formulation}) and the governing equations (Section~\ref{sec:governeqn}). We then describe the numerical algorithm for biofilm spreading (Section~\ref{sec:biofilm_numerical}), the simulation setup (Section~\ref{sec:simulation_setup}), and the grid convergence study (Section~\ref{sec:gridconvergence}). Section~\ref{sec:result1} analyzes the evolution of fracture permeability and biofilm dynamics. Section~\ref{sec:biofilmeffect} investigates the effects of biofilm growth on hydrogen concentration and identifies two distinct transport regimes, which are further characterized in Section~\ref{sec:chartype}. Finally, Section~\ref{sec:conclusion} provides the concluding remarks.

\section{Method}
\label{method}

\subsection{Problem formulation}
\label{sec:problem_formulation}
We consider flow and transport within a single fracture at the microscale.  
\begin{figure}[h!]
\centering
\includegraphics[width=14cm]{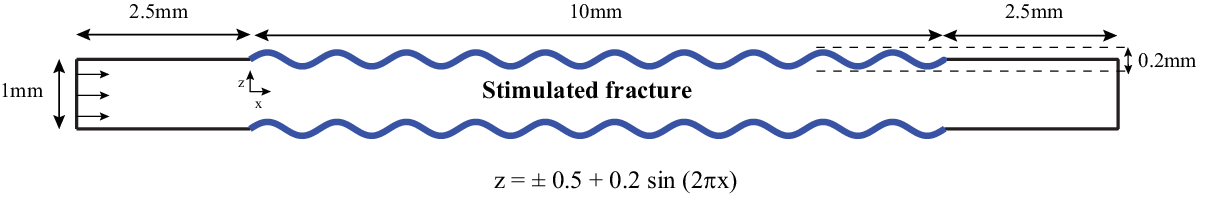}
\caption{Schematic for the computational domain and boundary conditions.}
\label{fig1}
\end{figure}
As shown in Fig. \ref{fig1}, we consider a rough fracture with olivine-bearing surfaces, where hydrogen is generated at the fluid-solid interface. The hydrogen generation rate is computed by fitting an exponential relationship between reaction rate ($r_{H_2}~[mol/m^2/s]$) and temperature ($T\ [K]$) using experimental data for $T < 568.15\ K$ \cite{mao2025geologic}. An empirical relationship is obtained as 

\begin{equation}
    r_{H_2} = 3.43\times 10^{-10} \exp (0.0167\ (T-273.15)). \label{eq:H2_generation}
\end{equation}

\noindent To represent the fracture roughness, the solid boundaries are prescribed as sinusoidal surfaces with an amplitude of $0.2\ mm$. At $T=315.15K$, the hydrogen generation flux $r_{H_2}$ is approximately $6.8\times10^{-10}~ mol/m^2/s$. Previous studies have suggested that hydrogen generation rates in stimulated geologic hydrogen systems may need to be enhanced by at least two orders of magnitude relative to natural rates to achieve economically viable production \cite{templeton2024low,yuan2026stimulated}. Here, we assume an approximately 520-fold enhancement in the hydrogen generation rate relative to the estimated natural serpentinization rate, corresponding to a surface-specific hydrogen generation rate of $r_{H_2} = 3.534\times10^{-7}~mol/m^{2}/s$.

In this system, microorganisms can attach to the fracture surfaces and form biofilms. Here, we consider sulfate reduction reaction mediated by sulfate reducing microorganisms (SRM), including species such as \textit{Oleidesulfovibrio alaskensis} and \textit{Archaeoglobus fulgidus} \cite{berta2018geochemical,liu2025growth,dopffel2025pressure,rosnes1991spore}. Sulfate-reducing microorganisms can remain active from near-ambient temperatures to approximately $70$--$90~^\circ$C, depending on the species. In such reactions, hydrogen serves as the electron donor, while sulfate is the electron acceptor and is reduced to sulfide. The reaction is given by

\begin{equation}
4 H_2+SO_4^{2-} +H^+ \rightarrow HS^{-}  +4 H_2O.
\label{eq:reaction}
\end{equation}

Microorganisms are represented as suspended bacteria transported by the fluid or as biofilms attached to the fracture surfaces. Biofilm growth modifies the local pore space and can substantially alter the fracture permeability. Suspended bacteria can attach to the solid surface and become biofilms, while biofilms can detach and return to the suspended phase. Biofilms also expand spatially through biomass growth and redistribution.

We further distinguish active SRM from inert biomass generated through microbial decay. Both active SRM and inert biomass can exist in the suspended phase or within the biofilm. We consider the following processes: (i) flow and mass transport in the free-flow fracture, (ii) flow and mass transport within the biofilm, (iii) growth and transport of suspended bacteria, and (iv) biofilm growth, decay, attachment, detachment, and spreading.

\subsection{Governing equations}
\label{sec:governeqn}
\subsubsection{Flow and reactive transport}
\label{sec:flow}
To simulate the coupled flow and mass transport in the free-flow and biofilm regions, we treat the biofilm as an evolving porous medium and solve the Darcy-Brinkman-Stokes (DBS) equations throughout the computational domain \cite{Soulaine2016,Vafai1981,Hsu1990,Bousquet2002,Goyeau2003,Kim2026-cw}. The local porosity, $\varepsilon_{\mathrm{f}}$, is related to the biofilm concentration as

\begin{gather}
\label{eq:vf}
\varepsilon_{\mathrm{f}}
=
\frac{\rho_{\mathrm{b}}^{0}}
{\rho_{\mathrm{b}}^{0}
+
\left(
\overline{X}_{SRM}^{\mathrm{b}}
+
\overline{X}_{in}^{\mathrm{b}}
\right)},
\end{gather}

\noindent where $\overline{X}_{SRM}^{\mathrm{b}}$ [\si{\kilogram\per\cubic\meter}] and $\overline{X}_{in}^{\mathrm{b}}$ [\si{\kilogram\per\cubic\meter}] are the volume-averaged biofilm concentrations of active SRM and inert biomass, respectively, and $\rho_{\mathrm{b}}^{0}$ [\si{\kilogram\per\cubic\meter}] is the dry biofilm density \cite{Kim2026-cw}. As the biofilm accumulates or decays, the local porosity changes, which further modifies the hydraulic properties of the biofilm.

The volume-averaged fluid velocity $\overline{\mathbf{u}}_{\mathrm{f}}$ and pore-fluid pressure $\overline{p}_{\mathrm{f}}$ are governed by

\begin{subequations}
\label{eq:dbs_eqs}
\begin{align}
\nabla\cdot\overline{\mathbf{u}}_{\mathrm{f}} &= 0,
\label{eq:dbs_continuity}\\
\frac{\rho_{\mathrm{f}}}{\varepsilon_{\mathrm{f}}}
\left[
\frac{\partial\overline{\mathbf{u}}_{\mathrm{f}}}{\partial t}
+
\nabla\cdot
\left(
\frac{1}{\varepsilon_{\mathrm{f}}}
\overline{\mathbf{u}}_{\mathrm{f}}
\overline{\mathbf{u}}_{\mathrm{f}}
\right)
\right]
&=
-\nabla\overline{p}_{\mathrm{f}}
+
\frac{\mu_{\mathrm{f}}}{\varepsilon_{\mathrm{f}}}
\nabla^{2}\overline{\mathbf{u}}_{\mathrm{f}}
-
\mu_{\mathrm{f}}k^{-1}\overline{\mathbf{u}}_{\mathrm{f}},
\label{eq:dbs_momentum}
\end{align}

\noindent where the volume-averaged fluid velocity $\overline{\mathbf{u}}_{\mathrm{f}}$ [\si{\meter\per\second}] and pore-fluid pressure $\overline{p}_{\mathrm{f}}$ [\si{\pascal}] are defined as

\begin{equation}
\overline{\mathbf{u}}_{\mathrm{f}} = \frac{1}{V}\int_{V_{\mathrm{f}}}\mathbf{u}_{\mathrm{f}}\,\dd V, \quad \overline{p}_{\mathrm{f}} = \frac{1}{V_{\mathrm{f}}}\int_{V_{\mathrm{f}}}p_{\mathrm{f}}\,\dd V.
\end{equation}
\end{subequations}

\noindent Here, $V$ [\si{\cubic\meter}] is the total averaging volume, $V_{\mathrm{f}}$ [\si{\cubic\meter}] is the pore-fluid volume, $\rho_{\mathrm{f}}$ [\si{\kilogram\per\cubic\meter}] and $\mu_{\mathrm{f}}$ [\si{\pascal\second}] are the density and dynamic viscosity of the fluid, respectively, and $k$ [\si{\square\meter}] is the local permeability. The last term in \Cref{eq:dbs_momentum} represents the flow resistance exerted by the biofilm. In the free-flow region ($\varepsilon_{\mathrm{f}}=1$), this resistance term vanishes and the equation reduces to the Navier-Stokes equation. As the biofilm accumulates ($\varepsilon_{\mathrm{f}} \rightarrow \varepsilon_{\mathrm{f,min}}$), the porous-medium resistance becomes significant, resulting in Darcy flow within the biofilm. The local permeability is calculated using the Kozeny-Carman relationship such that

\begin{equation}
\label{eq:KC}
k^{-1} = K_{0}^{-1} \frac{(1-\varepsilon_{\mathrm{f}})^2} {\varepsilon_{\mathrm{f}}^3},
\end{equation}

\noindent where $K_{0}$ [\si{\meter\squared}] is the intrinsic biofilm permeability.

Hydrogen and sulfate transport are modeled using advection–diffusion–reaction equations based on the sulfate-reduction reaction in \Cref{eq:reaction} such that 

\begin{subequations}
\label{eq:scalar_transport}
\begin{align}
\frac{\partial (\varepsilon_{\mathrm{f}}\overline{C}_{H})}{\partial t} + \nabla\cdot\left(\overline{\mathbf{u}}_{\mathrm{f}}\overline{C}_{H}\right) - \nabla\cdot\left(\varepsilon_{\mathrm{f}}D^{*}_{H}\nabla\overline{C}_{H}\right) &= -\alpha_{H}\left(\overline{X}_{SRM}^{\mathrm{sb}} + \overline{X}_{SRM}^{\mathrm{b}}\right)r_{SRM},
\label{eq:H2_transport}\\
\frac{\partial (\varepsilon_{\mathrm{f}}\overline{C}_{S})}{\partial t} + \nabla\cdot\left(\overline{\mathbf{u}}_{\mathrm{f}}\overline{C}_{S}\right) - \nabla\cdot\left(\varepsilon_{\mathrm{f}}D^{*}_{S}\nabla\overline{C}_{S}\right) &= -\left(\overline{X}_{SRM}^{\mathrm{sb}} + \overline{X}_{SRM}^{\mathrm{b}} \right)r_{SRM},
\label{eq:SO4_transport}
\end{align}
\end{subequations}

\noindent where $\overline{C}_{H}$ and $\overline{C}_{S}$ [\si{\kilogram\per\cubic\meter}] are the volume-averaged concentrations of hydrogen and sulfate, respectively, and $D^{*}_{H}$ and $D^{*}_{S}$ [\si{\square\meter\per\second}] are their corresponding effective diffusivities. $\overline{X}_{SRM}^{\mathrm{sb}}$ and $\overline{X}_{SRM}^{\mathrm{b}}$ [\si{\kilogram\per\cubic\meter}] are the suspended SRM concentration and SRM biofilm concentration, respectively. The stoichiometric coefficient $\alpha_H$ [\si{\kilogram\ce{H2}\per\kilogram\ce{SO4^{2-}}}] is determined from \Cref{eq:reaction} and is equal to $1/12$.


We describe the SRM-mediated reaction rate using multiplicative Monod kinetics as \cite{cassidy2015automated}

\begin{gather}
r_{SRM} = \hat{q}^{\max}_{SRM}\left(\frac{\overline{C}_{H}}{K_{SRM,H}+\overline{C}_{H}}\right)\left(\frac{\overline{C}_{S}}{K_{SRM,S}+\overline{C}_{S}}\right),
\end{gather}

\noindent where $\hat{q}^{\max}_{SRM}$ [\si{\per\second}] is the maximum sulfate utilization rate by SRM, and $K_{SRM,H}$ [\si{\kilogram\per\cubic\meter}] and $K_{SRM,S}$ [\si{\kilogram\per\cubic\meter}] are the half-saturation constants for hydrogen and sulfate, respectively.

\subsubsection{Suspended bacteria and biofilms}
\label{sec:bf}
We simulate both suspended bacteria and biofilms. In the suspended phase, active SRM and inert biomass are transported by advection and diffusion. Active SRM undergo growth and decay, while a fraction of the decayed biomass is converted to inert biomass. Both active SRM and inert biomass can attach to the biofilm or return to the suspended phase through detachment. Their suspended-phase concentrations are governed by

\begin{subequations}
\label{eq:biomass}
\begin{align}
\frac{\partial \left(\varepsilon_{\mathrm{f}}\overline{X}_{SRM}^{\mathrm{sb}}\right)}{\partial t}
&+
\nabla\cdot
\left(
\overline{\mathbf{u}}_{\mathrm{f}}
\overline{X}_{SRM}^{\mathrm{sb}}
\right)
-
\nabla\cdot
\left(
\varepsilon_{\mathrm{f}}D^{*}_{SRM}
\nabla\overline{X}_{SRM}^{\mathrm{sb}}
\right)
= \nonumber\\
& \quad \left(
Y_{SRM}r_{SRM}
-
b^{\mathrm{d}}_{SRM}
-
k^{\mathrm{att}}U^{\mathrm{att}}
\right)
\overline{X}_{SRM}^{\mathrm{sb}}
+
k^{\mathrm{det}}\tau^{\mathrm{det}}
\overline{X}_{SRM}^{\mathrm{b}},
\label{eq:suspended_biomass}
\\[10pt]
\frac{\partial \left(\varepsilon_{\mathrm{f}}\overline{X}_{in}^{\mathrm{sb}}\right)}{\partial t}
&+
\nabla\cdot
\left(
\overline{\mathbf{u}}_{\mathrm{f}}
\overline{X}_{in}^{\mathrm{sb}}
\right)
-
\nabla\cdot
\left(
\varepsilon_{\mathrm{f}}D^{*}_{in}
\nabla\overline{X}_{in}^{\mathrm{sb}}
\right)
= \nonumber\\
& \quad \beta_{in}b^{\mathrm{d}}_{SRM}
\overline{X}_{SRM}^{\mathrm{sb}}
-
k^{\mathrm{att}}U^{\mathrm{att}}
\overline{X}_{in}^{\mathrm{sb}}
+
k^{\mathrm{det}}\tau^{\mathrm{det}}
\overline{X}_{in}^{\mathrm{b}}.
\label{eq:suspended_biomass_inert}
\end{align}
\end{subequations}

For biofilms, the governing equations describe the local biomass growth and decay, whereas spatial redistribution is handled by the cellular automaton (CA) spreading algorithm described in \Cref{sec:biofilm_numerical}. The concentrations of active SRM and inert biomass in the biofilms are governed by

\begin{subequations}
\label{eq:biofilm_reactions}
\begin{align}
\dv{\left(\varepsilon_{\mathrm{f}}\overline{X}_{SRM}^{\mathrm{b}}\right)}{t}
&=
\left(
Y_{SRM}r_{SRM}
-
b^{\mathrm{d}}_{SRM}
-
k^{\mathrm{det}}\tau^{\mathrm{det}}
\right)
\overline{X}_{SRM}^{\mathrm{b}}
+
k^{\mathrm{att}}U^{\mathrm{att}}
\overline{X}_{SRM}^{\mathrm{sb}},
\label{eq:active_biofilm}\\
\dv{\left(\varepsilon_{\mathrm{f}}\overline{X}_{in}^{\mathrm{b}}\right)}{t}
&=
\beta_{in}b^{\mathrm{d}}_{SRM}
\overline{X}_{SRM}^{\mathrm{b}}
+
k^{\mathrm{att}}U^{\mathrm{att}}
\overline{X}_{in}^{\mathrm{sb}}
-
k^{\mathrm{det}}\tau^{\mathrm{det}}
\overline{X}_{in}^{\mathrm{b}}.
\label{eq:inactive_biofilm}
\end{align}
\end{subequations}

\noindent Here, $\overline{X}_{SRM}^{\mathrm{sb}}$ and $\overline{X}_{in}^{\mathrm{sb}}$ [\si{\kilogram\per\cubic\meter}] are the suspended-phase concentrations of active SRM and inert biomass, respectively, whereas $\overline{X}_{SRM}^{\mathrm{b}}$ and $\overline{X}_{in}^{\mathrm{b}}$ [\si{\kilogram\per\cubic\meter}] are their corresponding biofilm concentrations. $D^{*}_{SRM}$ and $D^{*}_{in}$ [\si{\square\meter\per\second}] are the effective diffusion coefficients for suspended SRM and inert biomass. The attachment process is characterized by the maximum attachment rate $k^{\mathrm{att}}$ [\si{\per\second}] and attachment probability $U^{\mathrm{att}}$ [-], while $k^{\mathrm{det}}$ [\si{\per\second}] and $\tau^{\mathrm{det}}$ [-] describe the maximum detachment rate and detachment probability, respectively. Microbial growth is coupled to sulfate utilization through the yield coefficient $Y_{SRM}$ [kg SRM/kg \ce{SO4^{2-}}], which is defined as the mass of active SRM produced per unit mass of sulfate consumed. Active SRM decay follows a first-order process with a decay coefficient $b^{\mathrm{d}}_{SRM}$ [\si{\per\second}], and a fraction $\beta_{in}=0.15$ [-] of the decayed active SRM is converted into inert biomass.

Suspended bacteria and biofilms are coupled through attachment and detachment. Previous studies have shown that microbial attachment depends on the local flow velocity \cite{Guanju2023,Minto2019}. Following \cite{Minto2019}, we compute $U^{\mathrm{att}}$ using a stepwise function of the local velocity magnitude ($U_{\mathrm{int}}$) as

\begin{align}
U_{\mathrm{att}} =
\begin{cases}
1, & \text{if } U_{\mathrm{int}} < U_{L},\\[4pt]
1 - \dfrac{\lvert U_{\mathrm{int}}\rvert-U_{L}}
{U_{U}-U_{L}},
& \text{if } U_{L}\leq U_{\mathrm{int}}\leq U_{U},\\[8pt]
0, & \text{if } U_{\mathrm{int}} > U_{U}.
\end{cases}
\end{align}

\noindent Here, $U_{\mathrm{int}}$ [\si{\meter\per\second}] is the local velocity magnitude, while $U_{L}$ and $U_{U}$ [\si{\meter\per\second}] are the lower and upper velocity thresholds for attachment, respectively.

Biofilm detachment is assumed to result from hydrodynamic shear acting on the biofilm. Shear-induced detachment occurs when the local shear stress exceeds a characteristic threshold \cite{Choi2003,Stoodley2001,Sudarsan2005}. Following \cite{Benioug2017}, we model $\tau^{\mathrm{det}}$ as a function of the local shear stress $\tau$ \si{\pascal} and the threshold shear stress $\tau_{\max}$ \si{\pascal} as

\begin{align}
\tau_{\mathrm{det}} =
\begin{cases}
0, & \text{if } \tau < \tau_{\max},\\[4pt]
\exp\!\left(
\dfrac{\tau-2\tau_{\max}}
{\tau_{\max}}
\right),
& \text{if } \tau_{\max}\leq\tau\leq2\tau_{\max},\\[8pt]
1, & \text{if } \tau\geq2\tau_{\max}.
\end{cases}
\end{align}

\noindent We apply a constant $\tau_{\max}$ for all simulations to isolate the effect of hydrodynamic forcing on biofilm accumulation.

\begin{figure}[h!]
    \centering
    \includegraphics[width=.75\textwidth]{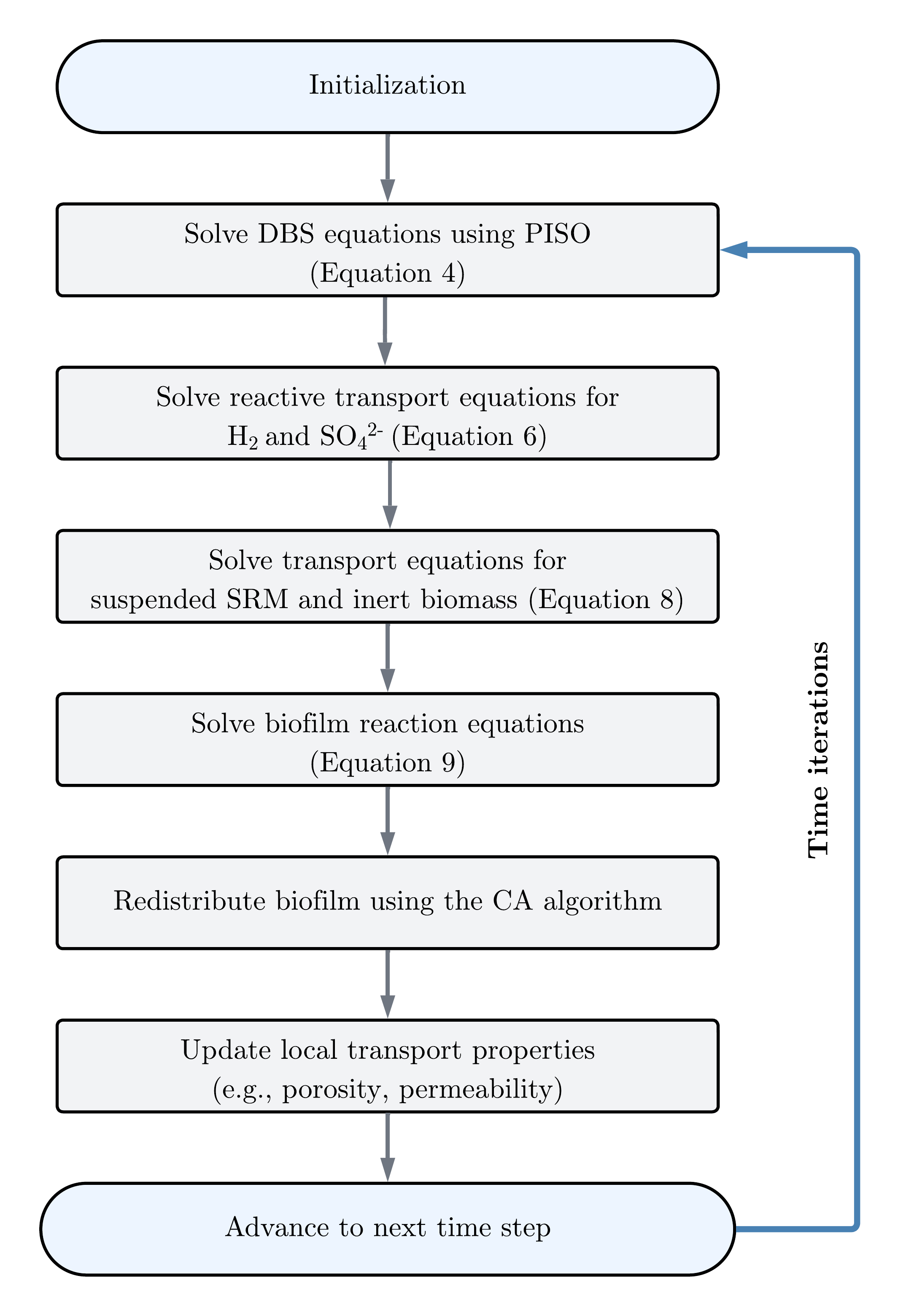}
    \caption{Numerical workflow for the coupled flow, reactive transport, and biofilm model. Biofilm growth and redistribution modify the local porosity and permeability, which are updated at each time step.}
    \label{fig_flowchart}
\end{figure}

\subsection{Biofilm spreading algorithm}
\label{sec:biofilm_numerical}

Local microbial growth can increase the biofilm concentration beyond the prescribed maximum value, $M_{\max}$. To capture the resulting biofilm expansion, we use a modified CA spreading algorithm based on previous approaches \cite{Picioreanu1998a,Tang2013}. The stochastic redistribution approach of \cite{Picioreanu1998a} can introduce artificial mixing among microbial populations and does not preserve the local biofilm composition. In contrast, the approach of \cite{Tang2013} preserves the spatial composition of the biofilm but requires a spreading path to be identified toward the biofilm-fluid interface, increasing computational complexity. In this study, we combine stochastic local spreading with conservative redistribution of individual biofilm populations. The local biofilm composition is therefore preserved without requiring a global search for the biofilm-fluid interface. The algorithm proceeds as follows:

\begin{description}

\item[Step 1.] \textbf{Identify source cells and excess biofilm.}
The local biofilm concentration is calculated as

\begin{align}
\overline{M}
=
\overline{X}_{SRM}^{\mathrm{b}}
+
\overline{X}_{in}^{\mathrm{b}},
\end{align}

\noindent with the maximum biofilm concentration $M_{\max}$ [\si{\kilogram\per\cubic\meter}] as

\begin{align}
M_{\max}
=
\rho_{\mathrm{b}}^{0}
\frac{1-\varepsilon_{\mathrm{f,min}}}
{\varepsilon_{\mathrm{f,min}}}
.
\end{align}

\noindent Here, cells with $\overline{M}>M_{\max}$ are identified as source cells for redistribution. For each source cell, the biofilm concentration transferred to neighboring cells is

\begin{align}
\mathcal{S}
=
\overline{M}_{\mathrm{src}}
-
\gamma M_{\max},
\end{align}

\noindent where $\gamma$ is the prescribed source-cell retention factor. When $\gamma=1$, $\mathcal{S}$ equals the excess biofilm concentration above $M_{\max}$. For $\gamma<1$, additional biofilm is redistributed such that the source-cell concentration is reduced to $\gamma M_{\max}$. A value of $\gamma=0.5$ was used in \cite{Picioreanu1998a}.

\item[Step 2.] \textbf{Classify neighboring cells.}
Face-connected neighboring cells with $\overline{M}\leq0.25M_{\max}$ are classified as low-biofilm-concentration cells, while the remaining neighboring cells are treated as partially occupied biofilm cells.

\item[Step 3.] \textbf{Select the receiving cell.}
If one or more low-biofilm-concentration neighboring cells are available, one is selected randomly as the receiving cell. Otherwise, a partially occupied neighboring cell is selected randomly. During successive redistribution, immediate transfer back to the previous source cell is avoided when alternative neighboring cells are available.

\item[Step 4.] \textbf{Transfer biofilm.}
The redistributed biofilm is partitioned between active SRM and inert biomass according to their local fractions in the source cell. For population $i$, where $i=\{SRM,in\}$,

\begin{align}
\phi_{i,\mathrm{src}}
&=
\frac{\overline{X}_{i,\mathrm{src}}^{\mathrm{b}}}
{\overline{M}_{\mathrm{src}}},
\\
\Delta M_i
&=
\mathcal{S}\phi_{i,\mathrm{src}}.
\end{align}

The biofilm concentrations in the source and receiving cells are then updated as

\begin{subequations}
\label{eq:CA_transfer}
\begin{align}
\widetilde{X}_{i,\mathrm{src}}^{\mathrm{b}}
&=
\overline{X}_{i,\mathrm{src}}^{\mathrm{b}}
-
\Delta M_i,
\label{eq:CA_source}\\
\widetilde{X}_{i,\mathrm{rec}}^{\mathrm{b}}
&=
\overline{X}_{i,\mathrm{rec}}^{\mathrm{b}}
+
\Delta M_i
\frac{V_{\mathrm{src}}}{V_{\mathrm{rec}}},
\label{eq:CA_receiver}
\end{align}
\end{subequations}

\noindent where ``src'' and ``rec'' refer to the source and receiving cells, respectively, $V_{\mathrm{src}}$ and $V_{\mathrm{rec}}$ [\si{\cubic\meter}] are their corresponding cell volumes, and the tilde indicates the biofilm concentration after redistribution. The volume ratio in \Cref{eq:CA_receiver} accounts for differences in cell volume during redistribution on a nonuniform mesh. Since each biofilm population is transferred according to its local fraction, the source-cell composition is retained during redistribution.

\item[Step 5.] \textbf{Repeat spreading.}
After redistribution, the total biofilm concentration in the receiving cell is recalculated. If it exceeds $M_{\max}$, the receiving cell becomes a new source cell and Steps 2-4 are repeated until no receiving cell exceeds $M_{\max}$.

\item[Step 6.] \textbf{Update transport properties.}
Following biofilm redistribution, the local fluid porosity and permeability are updated according to \Cref{eq:vf,eq:KC}. The updated fields are then used to solve the coupled flow and reactive transport equations at the next time step.

\end{description}

\noindent The coupled governing equations are implemented in OpenFOAM. At each time step, the DBS equations are first solved to obtain the pressure and velocity fields. The scalar transport equations are then solved for hydrogen, sulfate, and the suspended-phase concentrations of active SRM and inert biomass. Local biofilm reactions are subsequently evaluated, after which the CA algorithm redistributes biofilm when the local biofilm concentration exceeds $M_{\max}$. The porosity and permeability fields are then updated based on the redistributed biofilm and used in the next time step. Pressure-velocity coupling is achieved using the Pressure-Implicit with Splitting of Operators (PISO) algorithm \cite{ISSA1986}.

\subsection{Simulation setup and boundary conditions}
\label{sec:simulation_setup}

We prescribe a uniform inlet velocity, $\overline{\mathbf{u}}_{\mathrm{f}}=u_{\mathrm{in}}\mathbf{e}_{x}$, and fix the outlet pressure at $\overline{p}_{\mathrm{f}}=0$. A zero-gradient condition is applied to the outlet velocity, while no-slip conditions are imposed on both the serpentinizing and nonreactive walls. The inlet velocity is varied to obtain fracture Reynolds numbers of $Re_{\mathrm{f}}=120$, 160, and 240, where

\begin{align}
Re_{\mathrm{f}}
=
\frac{\rho_{\mathrm{f}}u_{\mathrm{in}}H}
{\mu_{\mathrm{f}}},
\end{align}

\noindent where $H$ [\si{\meter}] is the characteristic fracture aperture. The corresponding inlet velocities are $u_{\mathrm{in}}=0.12$, $0.16$, and $0.24$~\si{\meter\per\second}, respectively.

Hydrogen is produced only on the serpentinizing fracture walls. We impose the hydrogen generation flux as a Neumann boundary condition such that

\begin{align}
-\varepsilon_{\mathrm{f}}D^{*}_{H}
\nabla\overline{C}_{H}\cdot\mathbf{n}_{w}
=
r_{\mathrm{H_2}} MW_{\mathrm{H_2}},
\label{eq:H2_wall_bc}
\end{align}

\noindent where $r_{\mathrm{H_2}}$ [\si{\mole\per\square\meter\per\second}] is the prescribed hydrogen generation flux in \Cref{eq:H2_generation}, $MW_{\mathrm{H_2}}$ [\si{\kg\per\mole}] is the molecular weight of hydrogen, and $\mathbf{n}_{w}$ is the unit normal directed from the mineral surface into the fluid. At the inlet, we prescribe $\overline{C}_{H,\mathrm{in}}=0$~\si{\milli\gram\per\liter}, while zero-gradient conditions are applied at the outlet and on the nonreactive walls.

Sulfate is supplied through the inlet at a prescribed concentration of $\overline{C}_{S,\mathrm{in}}=\num{9.98e-2}$~\si{\kilogram\per\cubic\meter}. A zero-gradient condition is applied at the outlet. Since sulfate is neither generated nor consumed directly at the fracture surfaces, zero-flux boundary conditions are imposed on both reactive and nonreactive walls as

\begin{align}
\nabla\overline{C}_{S}\cdot\mathbf{n}_{w}=0.
\end{align}

For the suspended-phase concentrations of active SRM and inert biomass, and their corresponding biofilm concentrations, zero-gradient boundary conditions are applied at all boundaries. The complete boundary conditions are summarized in Table~\ref{tab:BC}.

\begin{table}[h!]
\centering
\caption{Boundary conditions used in the simulations.}
\label{tab:BC}
\resizebox{\textwidth}{!}{
\begin{tabular}{lllll}
\hline
Variable
& Inlet
& Outlet
& Serpentinizing wall
& Nonreactive wall \\
\hline

$\overline{\mathbf{u}}_{\mathrm{f}}$
& $\overline{\mathbf{u}}_{\mathrm{f}}
=u_{\mathrm{in}}\mathbf{e}_{x}$
& $\nabla\overline{\mathbf{u}}_{\mathrm{f}}\cdot\mathbf{n}=0$
& $\overline{\mathbf{u}}_{\mathrm{f}}=0$
& $\overline{\mathbf{u}}_{\mathrm{f}}=0$
\\[4pt]

$\overline{p}_{\mathrm{f}}$
& $\nabla\overline{p}_{\mathrm{f}}\cdot\mathbf{n}=0$
& $\overline{p}_{\mathrm{f}}=0$
& $\nabla\overline{p}_{\mathrm{f}}\cdot\mathbf{n}=0$
& $\nabla\overline{p}_{\mathrm{f}}\cdot\mathbf{n}=0$
\\[4pt]

$\overline{C}_{H}$
& $\overline{C}_{H}=0$
& $\nabla\overline{C}_{H}\cdot\mathbf{n}=0$
& $-\varepsilon_{\mathrm{f}}D^{*}_{H}
\nabla\overline{C}_{H}\cdot\mathbf{n}_{w}
=r_{\mathrm{H_2}} MW_{\mathrm{H_2}}$
& $\nabla\overline{C}_{H}\cdot\mathbf{n}=0$
\\[4pt]

$\overline{C}_{S}$
& $\overline{C}_{S}=\overline{C}_{S,\mathrm{in}}$
& $\nabla\overline{C}_{S}\cdot\mathbf{n}=0$
& $-\varepsilon_{\mathrm{f}}D^{*}_{S}
\nabla\overline{C}_{S}\cdot\mathbf{n}_{w}=0$
& $-\varepsilon_{\mathrm{f}}D^{*}_{S}
\nabla\overline{C}_{S}\cdot\mathbf{n}_{w}=0$
\\[4pt]

$\overline{X}_{i}^{\mathrm{sb}}$,
$i=\{SRM,in\}$
& $\nabla\overline{X}_{i}^{\mathrm{sb}}\cdot\mathbf{n}=0$
& $\nabla\overline{X}_{i}^{\mathrm{sb}}\cdot\mathbf{n}=0$
& $\nabla\overline{X}_{i}^{\mathrm{sb}}\cdot\mathbf{n}=0$
& $\nabla\overline{X}_{i}^{\mathrm{sb}}\cdot\mathbf{n}=0$
\\[4pt]

$\overline{X}_{i}^{\mathrm{b}}$,
$i=\{SRM,in\}$
& $\nabla\overline{X}_{i}^{\mathrm{b}}\cdot\mathbf{n}=0$
& $\nabla\overline{X}_{i}^{\mathrm{b}}\cdot\mathbf{n}=0$
& $\nabla\overline{X}_{i}^{\mathrm{b}}\cdot\mathbf{n}=0$
& $\nabla\overline{X}_{i}^{\mathrm{b}}\cdot\mathbf{n}=0$
\\

\hline
\end{tabular}
}
\end{table}


Initially, hydrogen, sulfate, suspended active SRM, suspended inert biomass, and inert biofilm concentrations are set to zero throughout the computational domain. We initialize the active SRM biofilm only in the first computational cell adjacent to the serpentinizing fracture wall, where the initial biofilm concentration is prescribed as $\overline{X}_{SRM}^{\mathrm{b}}(t=0)=5$~\si{\kilogram\per\cubic\meter}. All remaining cells are initialized with zero active SRM biofilm concentration.

To quantify the effect of biofilm transport properties, we systematically vary the intrinsic biofilm permeability such that

\begin{align}
K_0 =
2\times10^{-10},\;
1\times10^{-10},\;
1\times10^{-11},\;
1\times10^{-12},\;
1\times10^{-13}
\ \si{\meter\squared}.
\end{align}

Combining these conditions with the three fracture Reynolds numbers yields 15 simulation cases. The threshold shear stress $\tau_{\max}$ and microbial kinetic parameters are kept constant across all cases to isolate the effects of fracture flow and intrinsic biofilm permeability.

All simulations are performed for approximately $1\times10^{6}$ s, corresponding to approximately 12 days of biofilm development. The physical, transport, and microbial parameters used in the simulations are summarized in Table~\ref{tab:parameters}.

\begin{table}[h!]
\centering
\caption{Physical, transport, and microbial parameters used in the simulations.}
\label{tab:parameters}
\begin{tabular}{llll}
\hline
Parameter & Symbol & Value & Unit \\
\hline

Fluid density
& $\rho_{\mathrm{f}}$
& $1000$
& \si{\kilogram\per\cubic\meter}
\\

Dynamic viscosity
& $\mu_{\mathrm{f}}$
& $1.0\times10^{-3}$
& \si{\pascal\second}
\\

Hydrogen effective diffusivity
& $D^{*}_{H}$
& $6.7\times10^{-9}$
& \si{\square\meter\per\second}
\\

Sulfate effective diffusivity
& $D^{*}_{S}$
& $1.07\times10^{-9}$
& \si{\square\meter\per\second}
\\

SRM effective diffusivity
& $D^{*}_{SRM}$
& $1.00\times10^{-12}$
& \si{\square\meter\per\second}
\\

Inert biomass effective diffusivity
& $D^{*}_{in}$
& $1.00\times10^{-12}$
& \si{\square\meter\per\second}
\\

Dry biofilm density
& $\rho_{\mathrm{b}}^{0}$
& $60$
& \si{\kilogram\per\cubic\meter}
\\

Maximum biofilm concentration
& $M_{\max}$
& $27.7$
& --
\\

Minimum fluid porosity
& $\varepsilon_{\mathrm{f,min}}$
& $0.7$
& --
\\

Maximum sulfate utilization rate
& $\hat{q}^{\max}_{SRM}$
& $2.64\times10^{-3}$
& \si{\per\second}
\\

Hydrogen half-saturation constant
& $K_{SRM,H}$
& $4.80\times10^{-6}$
& \si{\kilogram\per\cubic\meter}
\\

Sulfate half-saturation constant
& $K_{SRM,S}$
& $2.88\times10^{-3}$
& \si{\kilogram\per\cubic\meter}
\\

SRM yield coefficient
& $Y_{SRM}$
& $0.05$
& \si{\kilogram\per\kilogram}
\\

SRM decay coefficient
& $b^{\mathrm{d}}_{SRM}$
& $1.20\times10^{-6}$
& \si{\per\second}
\\

Decay-to-inert fraction
& $\beta_{in}$
& $0.15$
& --
\\

Maximum attachment rate
& $k^{\mathrm{att}}$
& $0.1$
& \si{\per\second}
\\

Lower attachment velocity
& $U_L$
& $1.00\times10^{-12}$ 
& \si{\meter\per\second}
\\

Upper attachment velocity
& $U_U$
& $0.0003$
& \si{\meter\per\second}
\\

Maximum detachment rate
& $k^{\mathrm{det}}$
& $0.1$
& \si{\per\second}
\\

Threshold shear stress
& $\tau_{\max}$
& $1.39$
& \si{\pascal}
\\

CA retention factor
& $\gamma$
& $1$
& --
\\

Sulfate inlet concentration
& $\overline{C}_{S,\mathrm{in}}$
& $9.98\times10^{-2}$
& \si{\kilogram\per\cubic\meter}
\\

\hline
\end{tabular}
\end{table}

\subsection{Grid convergence}
\label{sec:gridconvergence}
To assess grid sensitivity, we conducted a grid-convergence study by successively refining the computational mesh. Simulations were performed using grids with average cell sizes of $0.2$, $0.1$, $0.05$, $0.025$, and $0.0125~\mathrm{mm}$. In the $z$-direction, the mesh was refined near the walls using the \texttt{simpleGrading} function in OpenFOAM with a grading factor of 25, providing finer resolution near the reactive surfaces.

\begin{figure}[htbp]
\centering
\includegraphics[width=14cm]{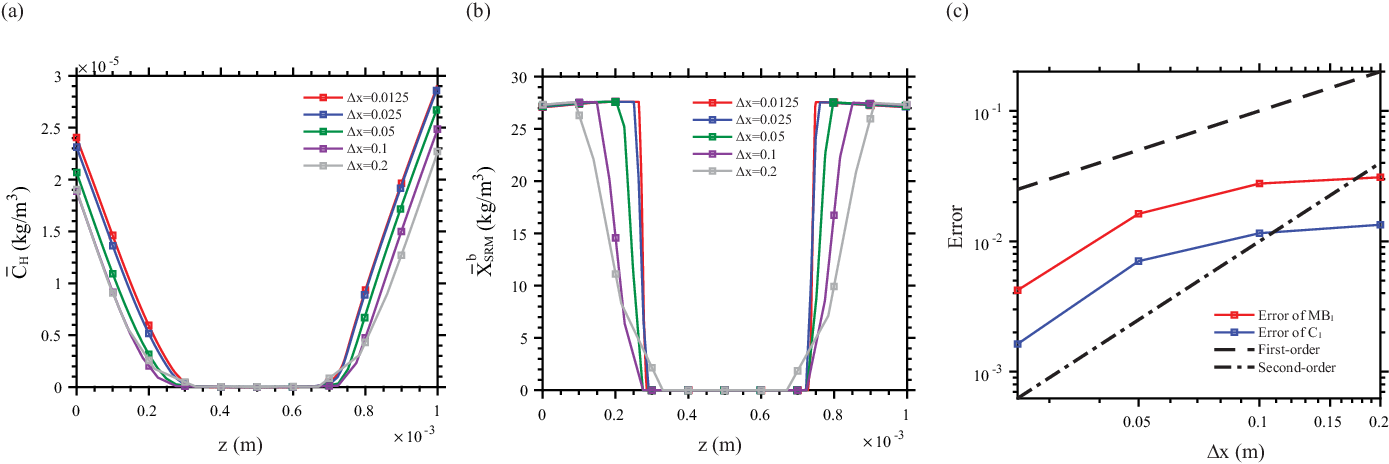}
\caption{Grid-convergence analysis showing (a) hydrogen concentration $\overline{C}_{H}$, (b) active SRM biofilm concentration $\overline{X}_{SRM}^{\mathrm{b}}$, and (c) relative $L_2$ errors for different grid sizes.}
\label{fig2}
\end{figure}

The grid-convergence simulations were performed at $K_0=10^{-13}$~\si{\meter\squared} and $u_{\mathrm{in}}=0.16$~\si{\meter\per\second}. Fig.~\ref{fig2}(a)--(b) show the profiles of hydrogen concentration and active SRM biofilm concentration along the $z$-direction at $x=5$~\si{\milli\meter}. For average grid sizes of $0.1~\mathrm{mm}$ and below, the hydrogen and active SRM biofilm concentration profiles show close agreement, indicating that the solutions approach grid independence. Taking the simulation results obtained with $\Delta x=0.0125~\mathrm{mm}$ as the reference solution, the 2-norm relative errors between the profiles were calculated after interpolating each solution onto 100 uniformly distributed points along the $z$-axis.

The relative errors are shown in Fig.~\ref{fig2}(c) on a logarithmic scale. The two dashed lines show the reference slopes corresponding to first-order convergence ($O(\Delta x)$) and second-order convergence ($O((\Delta x)^2)$). For the finer grids, the errors decrease at a rate close to second-order convergence. Based on the agreement between concentration profiles and the observed error convergence, we selected a grid size of 0.025~\si{\milli\meter} for the subsequent simulations.

\section{Results and Discussion}

\subsection{Evolution of fracture permeability and biofilm dynamics}\label{sec:result1}
\begin{figure}[htbp]
\centering
\includegraphics[width=14cm]{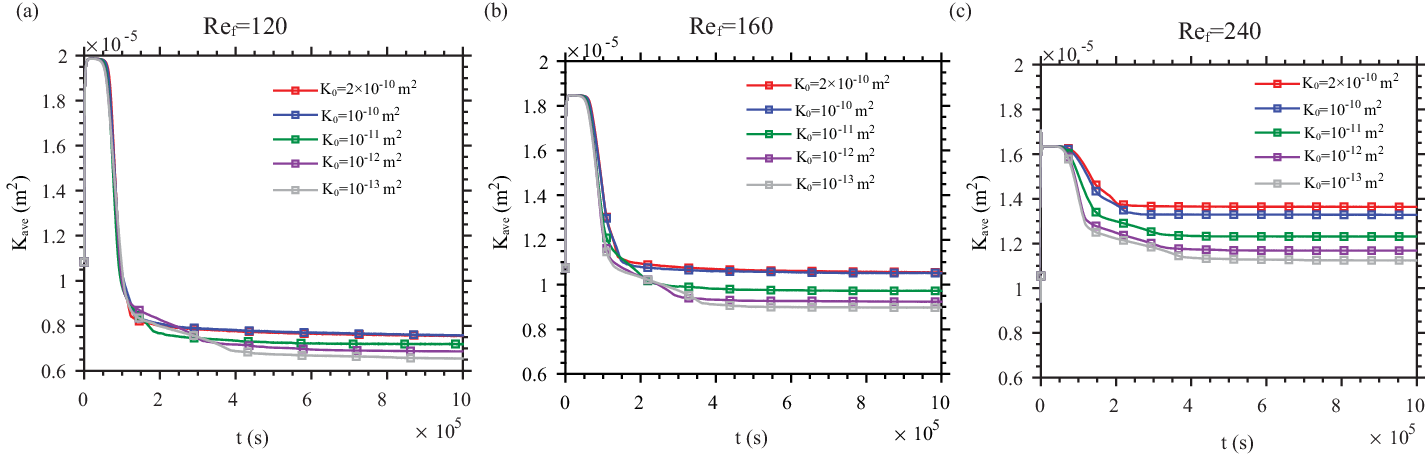}
\caption{Evolution of apparent average fracture permeability at} (a) $Re_f=120$, (b) $Re_f=160$, and (c) $Re_f=240$.
\label{fig3}
\end{figure}

\begin{figure}[htbp]
\centering
\includegraphics[width=14cm]{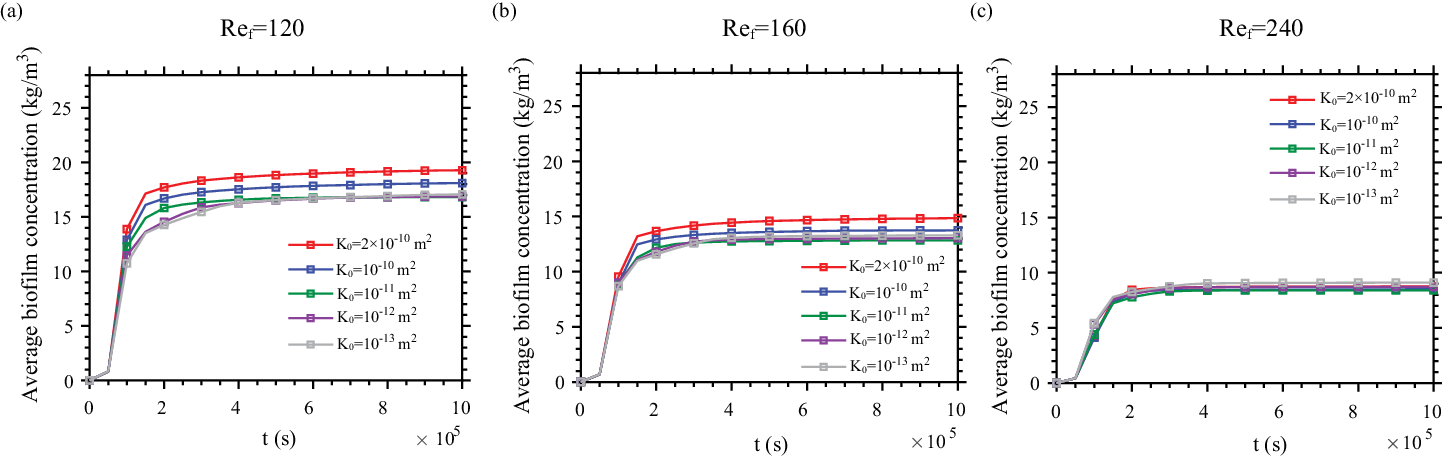}
\caption{Evolution of average biofilm concentration a (a) $Re_f=120$, (b) $Re_f=160$, and (c) $Re_f=240$.}
\label{fig4}
\end{figure}

Biofilm growth modifies the hydraulic properties of the fracture by reducing the available flow aperture and redistributing the local velocity field. In this system, the apparent average fracture permeability is defined as
\begin{equation}
K_{\mathrm{ave}} = \frac{\mu u L}{p_{\mathrm{in}}-p_{\mathrm{out}}},
\end{equation}
where $\mu~[Pa\cdot s]$ is the dynamic viscosity of water, $u~[m/s]$ is the average flow velocity, $L~[m]$ is the fracture length, and $p_{\mathrm{in}}-p_{\mathrm{out}}~[Pa]$ is the pressure drop across the fracture. Fig.~\ref{fig3} shows the evolution of $K_{\mathrm{ave}}$ for different fracture Reynolds numbers ($Re_f$) and intrinsic biofilm permeabilities ($K_0$). As biofilm accumulates along the fracture walls, the effective fracture aperture decreases, reducing $K_{\mathrm{ave}}$. The permeability eventually approaches a quasi-steady state, as changes in biofilm thickness and morphology stabilize. This behavior is consistent with the biofilm accumulation shown in Fig.~\ref{fig4}, which reflects the competing effects of microbial growth, reactant transport, and hydrodynamic shear. The initial apparent permeability, $K_{\mathrm{ave}}(t=0)$, also decreases with increasing $Re_f$, even before biofilm growth. This dependence reflects the nonlinear relationship between pressure drop and flow rate in rough fractures associated with inertial effects, consistent with literature \cite{chen2015evaluation}.

Across the range of flow rates considered, biofilm growth reduces the apparent average fracture permeability by up to $65\%$. The reduction is less significant at higher $Re_f$, where stronger hydrodynamic shear limits biofilm accumulation, resulting in less accumulated biomass and a thinner biofilm layer (Fig.~\ref{fig4}). Here, the total biofilm mass is calculated by integrating the local biofilm biomass concentration over the fracture domain. The effect of flow rate on biofilm accumulation is also evident in the biofilm morphologies shown in Fig.~\ref{fig6}.

\begin{figure}[htbp]
\centering
\includegraphics[width=14cm]{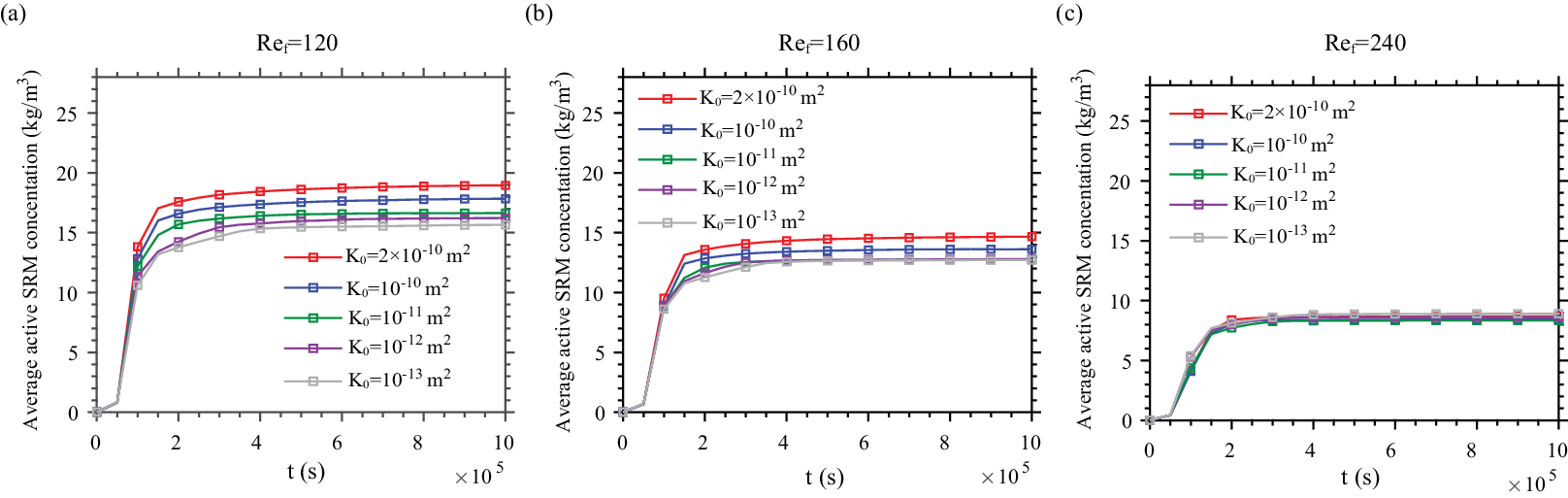}
\caption{Evolution of average active SRM biofilm concentration at (a) $Re_f=120$, (b) $Re_f=160$, and (c) $Re_f=240$.}
\label{fig11}
\end{figure}

For a given flow rate, intrinsic biofilm permeability also affects the rate and extent of biofilm accumulation. Higher biofilm permeability facilitates reactant transport within the biofilm matrix, leading to faster growth and greater biomass accumulation (Fig.~\ref{fig4}). This effect is particularly relevant in hydrogen-generating fractures, where hydrogen is produced at the fracture wall while sulfate ($SO_4^{2-}$) is supplied from the bulk fluid. The two reactants are therefore transported from opposite sides of the biofilm, and lower biofilm permeability can restrict their availability to active microorganisms. As biofilm accumulates, it reduces the local fracture aperture and increase the hydrodynamic shear. This explains the higher biofilm mass at lower $Re_f$. At higher $Re_f$, the thinner biofilm reduces internal mass transport limitations, diminishing the influence of intrinsic biofilm permeability on active biomass and resulting in the curves in Fig.~\ref{fig11}(c) to nearly overlap. This two-way coupling between flow and biofilm dynamics controls the evolution of fracture permeability and ultimately affects the produced hydrogen concentration.

The evolution of total biofilm accumulation does not fully describe microbial activity, as the biofilm contains both active and inert biomass. To distinguish their contributions, we examine the evolution of total active biofilm concentration under different $Re_f$ and intrinsic biofilm permeabilities (Fig.~\ref{fig11}). For highly permeable biofilms, the active biomass continues to increase before approaching a quasi-steady state, consistent with greater reactant accessibility within the biofilm. In contrast, the active biomass remains nearly constant for low-permeability biofilms (Fig.~\ref{fig11}a--c), despite stronger internal transport limitations. This behavior may also be attributed to the relatively high inlet sulfate concentration, reducing the sensitivity of microbial growth to mass transport limitations. At lower sulfate concentrations, transport limitations may become more pronounced, leading to a reduction in active biomass over time.

\begin{figure}[htbp]
\centering
\includegraphics[width=14cm]{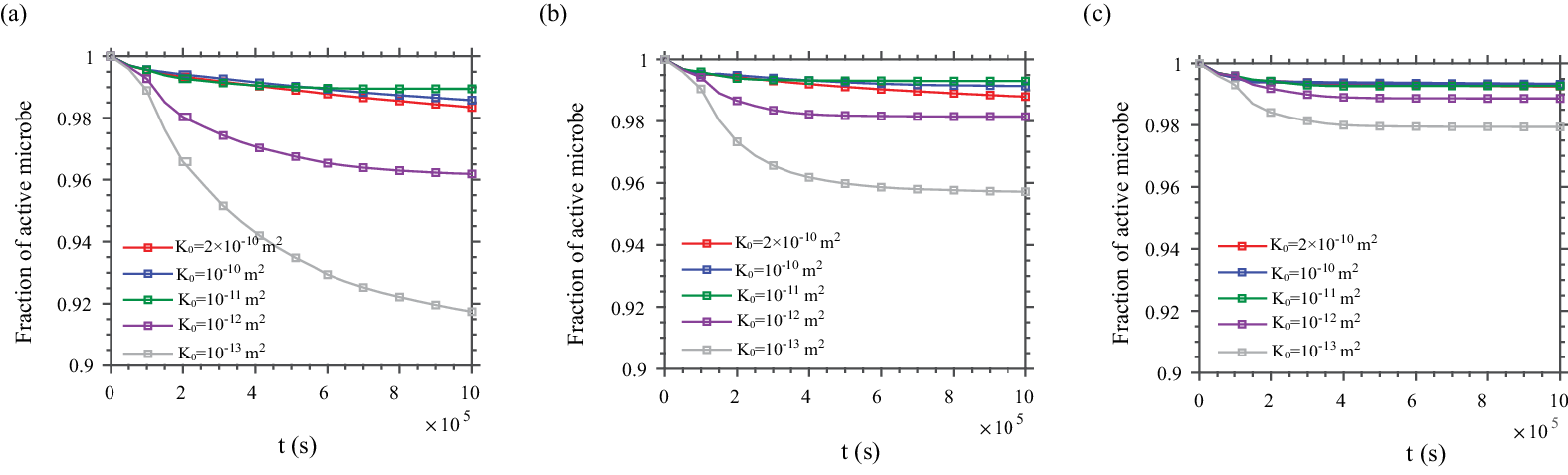}
\caption{Evolution of active biofilm mass fraction at (a) $Re_f=120$, (b) $Re_f=160$, and (c) $Re_f=240$}
\label{fig12}
\end{figure}

To distinguish changes in biofilm composition from total biomass accumulation, we define the active biofilm mass fraction $\eta$ as 
\begin{equation}
\eta(t) =
\frac{\displaystyle\sum_{i=1}^{N}
\overline{X}_{SRM,i}^{\mathrm{b}}(t)\,V_i}
{\displaystyle\sum_{i=1}^{N}
\overline{M}_i(t)\,V_i},
\end{equation}
where $V_i$ is the volume of computational cell $i$ and $N$ is the total number of cells. Fig.~\ref{fig12} shows the evolution of $\eta$, which decreases with time for all three fracture Reynolds numbers. This trend indicates that inert biomass progressively accumulates relative to active biomass as the biofilm develops. As biofilm thickness increases, longer transport distance between the bulk fluid and the reactive wall may limit reactant availability within the biofilm, resulting in biomass decay. 

The magnitude of this decrease in $\eta$ depends on both biofilm permeability and flow rate. Higher biofilm permeability generally maintains a higher active biomass fraction, consistent with more rapid reactant transport within biofilm. Similarly, higher $Re_f$ enhances reactant transport through advection, while stronger hydrodynamic shear limits biofilm accumulation and maintains a thinner biofilm layer. These effects favor microbial growth by improving reactant accessibility and reducing the accumulation of inert biomass compared with low-flow conditions. As a result, higher flow rates and intrinsic biofilm permeability tend to sustain a larger active biomass fraction, even when the total biofilm accumulation is reduced.

The distinction between total and active biomass is important for understanding the effects of biofilm growth on hydrogen transport. While total biofilm accumulation controls the reduction in available flow aperture, active biomass determines the potential for microbial hydrogen consumption. Biofilm permeability and fracture flow therefore influence hydrogen recovery through their effects on both transport resistance and microbial activity. Since intrinsic biofilm permeability is difficult to measure directly, the combined evolution of total biofilm mass and active biomass fraction ($\eta$) under different $Re_f$ may provide useful constraints for parameterizing biofilm permeability. In particular, the dependence of $\eta$ on flow rate and biofilm permeability suggests that biofilm population dynamics may serve as an indirect indicator of the effective transport properties of the biofilm. These coupled changes provide the basis for interpreting the outlet hydrogen concentration evolution discussed in the following section.

\subsection{Effect of biofilm permeability on hydrogen transport}
\label{sec:biofilmeffect}

This section examines how hydrogen transport is impacted by fracture flow and biofilm dynamics. In the serpentinizing fracture considered here, hydrogen generated at the reactive fracture wall must be transported through or around the biofilm before entering the bulk fluid. During this process, hydrogen can be transported by advection and diffusion within the porous biofilm matrix while being consumed by active microorganisms. Biofilm growth therefore affects the produced hydrogen concentration through both transport resistance and microbial consumption. As the biofilm thickens, the increasing transport distance between the reactive wall and the bulk fluid would generally be expected to reduce the outlet hydrogen concentration. Our simulations, however, reveal two distinct patterns of outlet hydrogen concentration evolution, depending on intrinsic biofilm permeability.

\begin{figure}[htbp]
\centering
\includegraphics[width=14cm]{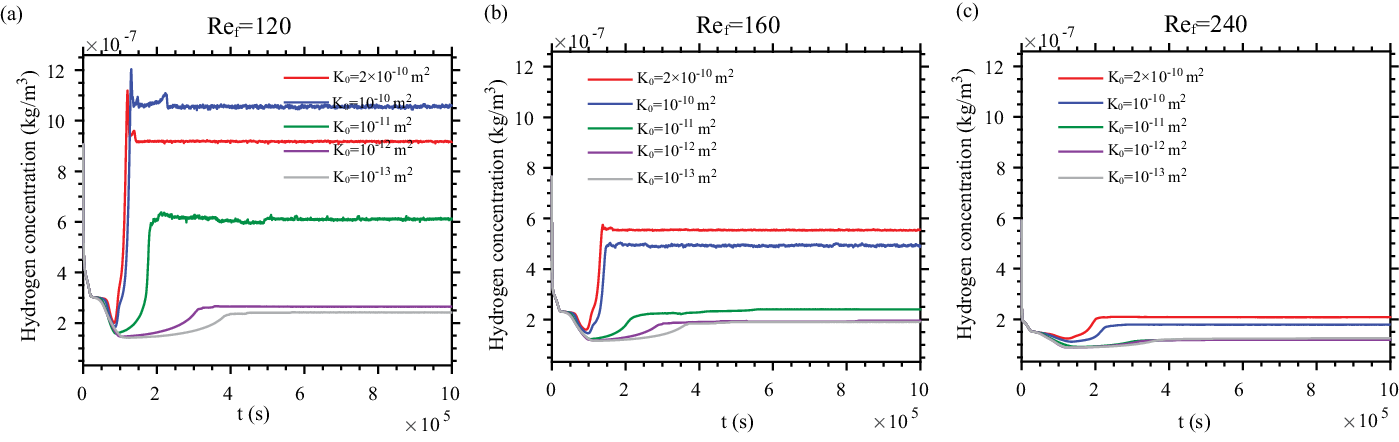}
\caption{Temporal evolution of outlet hydrogen concentration for (a) $Re_f=120$, (b) $Re_f=160$, and (c) $Re_f=240$. This 2D simulation assumes z direction is $2mm$.}
\label{fig5}
\end{figure}

Fig.~\ref{fig5} shows the temporal evolution of outlet hydrogen concentration for different fracture Reynolds numbers ($Re_f$) and intrinsic biofilm permeabilities ($K_0$). For low-permeability biofilms ($K_0=1\times10^{-12}$ and $1\times10^{-13}~\mathrm{m^2}$), $\overline{C}_{H}$ decreases approximately monotonically as the biofilm develops, followed by a slight increase at late times. This increase coincides with a reduction in active biomass (Fig.~\ref{fig11}), suggesting a decrease in microbial hydrogen consumption. In contrast, $\overline{C}_{H}$ exhibits a non-monotonic trend for highly permeable biofilms ($K_0=2\times10^{-10}, 1\times10^{-10}~\mathrm{m^2}, 1\times10^{-11}~\mathrm{m^2}$). For example, at $K_0=2\times10^{-10}~\mathrm{m^2}$ and $Re_f = 120$, $\overline{C}_{H}$ initially decreases until approximately $t=1.0\times10^5~s$ and then increases sharply until $t=1.5\times10^5~s$, and finally decreases towards a quasi-steady state. Similar temporal responses are observed
at all three fracture Reynolds numbers simulated (Fig.~\ref{fig5})(a-c). We classify these two patterns as Type I and Type II, corresponding to non-monotonic hydrogen concentration evolution in highly permeable biofilms and an overall decreasing $\overline{C}_{H}$ in low-permeability biofilms, respectively.

\begin{figure}[htbp]
\centering
\includegraphics[width=14cm]{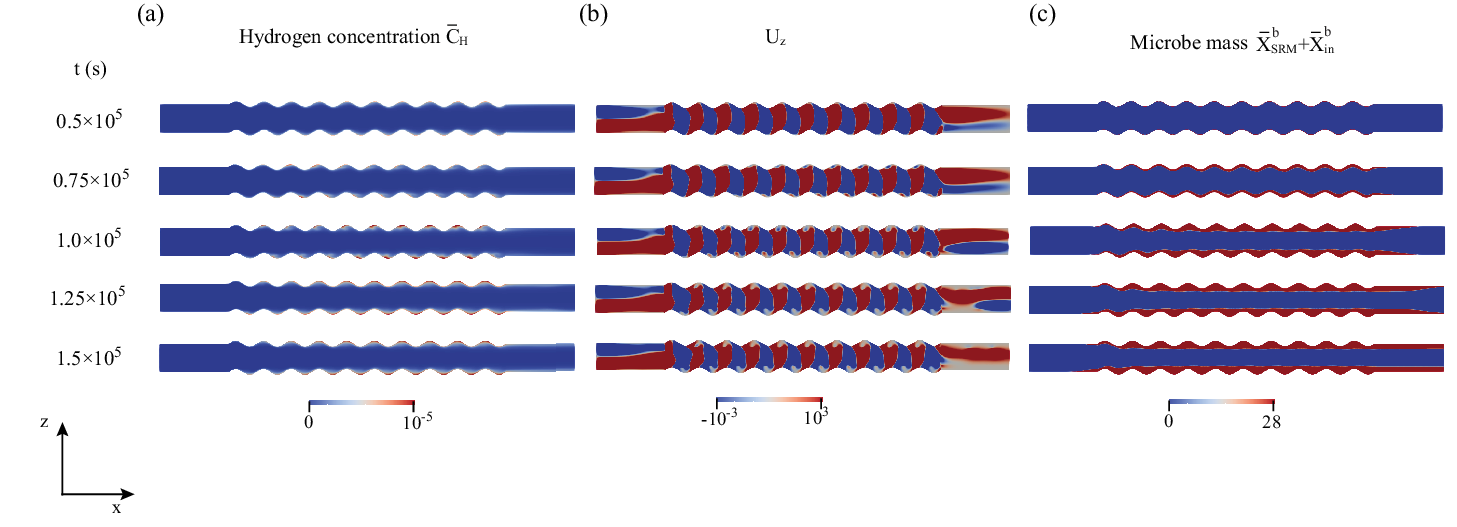}
\caption{Temporal evolution of (a) hydrogen concentration $C_H$, (b) transverse velocity, and (c) biofilm concentration ($\overline{X}_{SRM}^{\mathrm{b}}+ \overline{X}_{in}^{\mathrm{b}}$) for $Re_f=160$ and $K_0=2\times10^{-10}~\mathrm{m^2}$.}
\label{fig6}
\end{figure}

The difference between Type I and Type II responses is associated with changes in the local flow field as biofilms develop within the fracture. Figure~\ref{fig6} illustrates the  hydrogen concentration fields, flow fields, and biofilm distribution for a representative Type I case with $Re_f=160$ and $K_0=2\times10^{-10}~m^2$. As shown in Fig.~\ref{fig6}(a), hydrogen accumulates within the biofilm and near the reactive wall between $t=0.5\times10^5~s$ and $t=1.0\times10^5~s$. The latter time approximately coincides with the minimum outlet hydrogen concentration. During this stage, the local flow structure may restrict hydrogen transport from the near-wall region into the bulk fracture fluid. As shown in Fig.~\ref{fig6}(b), the flow field exhibits a vortex structure that reduces hydrogen advection into the bulk fluid. Similar eddy structures have been reported in other studies on flow through rough fractures in \cite{briggs2017numerical,wang2020complexity,lee2015tail}. As hydrogen continues to be generated at the reactive wall while advection exchange is restricted, the local hydrogen concentration increases during this period. Local high hydrogen concentration promotes biofilm growth during this stage as shown in Fig. \ref{fig4}(a). 

As the biofilm grows, it modifies the local pore space and suppresses the roughness-induced flow structure. Although the biofilm is highly permeable in the Type I cases, its permeability is still much lower than that of the open fracture. Therefore, biofilm accumulation reduces the local velocity near the wall and weakens the eddy structure. This transition is evident from Fig.~\ref{fig6}(b) and (c). Under the same color scale, the near-wall eddy diminishes from $t=1.0\times10^5~s$. By $t=1.5\times10^5~s$, the biofilm has accumulated sufficiently to fill the fracture-wall roughness. Once this local flow structure disappears, the previously trapped hydrogen is transported more effectively into the bulk fracture fluid. As a result, the outlet hydrogen concentration begins to increase at approximately $t=1.0\times10^5~s$, leading to the transient $\overline{C}_{H}$ enhancement characteristic as observed in Type I.

\begin{figure}[htbp]
\centering
\includegraphics[width=14cm]{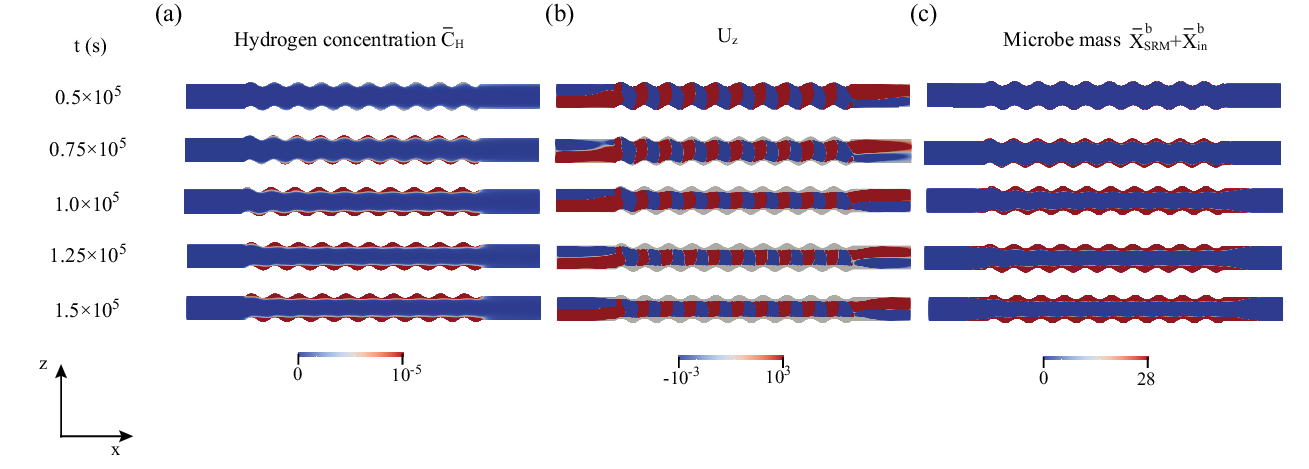}
\caption{Temporal evolution of (a) hydrogen concentration $C_H$, (b) transverse velocity, and (c) biofilm concentration ($\overline{X}_{SRM}^{\mathrm{b}}+ \overline{X}_{in}^{\mathrm{b}}$) for $Re_f=160$ and $K_0=10^{-13}~\mathrm{m^2}$.}
\label{fig7}
\end{figure}

The evolution of Type II is significantly different because hydrogen transport is primarily limited by the low permeability of the biofilm matrix. Fig~\ref{fig7} shows a representative case with $Re_f=160$ and $K_0=1\times10^{-13}~\mathrm{m^2}$. Although near-wall recirculation is initially present, the formation of even a thin biofilm layer restricts advective transport near the reactive wall. Unlike Type I, the influence of roughness-induced flow structures becomes secondary to the transport resistance of the biofilm. Hydrogen generated at the wall is therefore transported increasingly by diffusion through the biofilm, where it is also consumed by active microorganisms. As biofilm thickness increases, the longer diffusion distance and continued microbial consumption are associated with a progressive decrease in outlet hydrogen concentration.

Once biofilm growth approaches a quasi-steady state, the hydrogen concentration in Type II cases exhibits only a slight recovery. This late-time behavior coincides with the decreasing active biomass fraction (Fig.~\ref{fig12}), suggesting a changing balance between microbial consumption and hydrogen transport as the biofilm matures. The overall outlet $\overline{C}_{H}$ evolution, however, remains dominated by the transport limitation imposed by the low-permeability biofilm. The distinction between Type I and Type II, therefore, lies primarily in whether changes in local fracture flow or transport resistance within the biofilm govern the transient hydrogen concentration. The overall response is dominated by the transport limitation imposed by the low-permeability biofilm at early times, while the long-tail response is due to microbe dynamics.

This transition between Type I and Type II can be described by the Darcy-scale Reynolds number 

\begin{equation}
    Re_D  = \frac{u \sqrt{K_0}}{\nu}
\end{equation}

\noindent where $K_0$ is the intrinsic biofilm permeability and $\nu~[m^2]$ is the water kinematic viscosity. $Re_D$ depicts the confinement effect of biofilm matrix on the free-flow behavior. The parameter $Re_D$ provides a measure of the extent to which flow can penetrate the permeable biofilm matrix. When $Re_D$ is relatively large, advection through the biofilm can influence near-wall hydrogen transport, allowing local flow structures and their subsequent suppression by biofilm growth to affect the outlet $\overline{C}_{H}$. This corresponds to the Type I regime. When $Re_D$ is small, the biofilm strongly confines the flow, and hydrogen transport across the biofilm is dominated by diffusion, making the impact from flow structure negligible (Type II).

\begin{figure}[htbp]
\centering
\includegraphics[width=7cm]{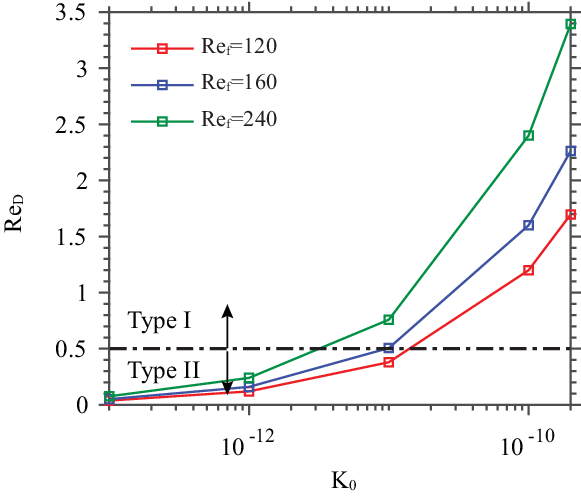}
\caption{Classification of Type I and Type II hydrogen transport regimes in the $K_0$--$Re_D$ parameter space.}
\label{fig8}
\end{figure}

Based on the simulation cases examined in this study, the transition between Type I and Type II behavior occurs at an approximate threshold of $Re_D \approx 0.5$. However, this threshold should not be considered universal, as its applicability across different conditions remains to be established. Fig.~\ref{fig8} summarizes the simulation results in the $K_0$--$Re_D$ parameter space, where the upper and lower regions correspond to Type II and Type I behavior, respectively. This transition highlights the substantial influence of intrinsic biofilm properties on hydrogen transport and apparent hydrogen production in stimulated geologic hydrogen systems. This distinct transport behavior associated with Type I and Type II regimes may provide a practical means of constraining or parameterizing biofilm permeability, as this property is difficult to measure directly under in situ conditions.

\subsection{Characteristic hydrogen concentration responses in Type I and Type II regime}
\label{sec:chartype}
Section \ref{sec:biofilmeffect} identifies two distinct types of hydrogen transport behavior and discusses their controlling parameters. Here, we further characterize these two types using the minimum and maximum quasi-steady state outlet hydrogen concentrations $\overline{C}_{H}$ and the time required to approach a quasi-steady state. The minimum and maximum hydrogen concentrations constrain production forecasts and uncertainty quantification for stimulated geologic hydrogen systems. The timescale to reach a quasi-steady state indicates when the produced hydrogen concentration is expected to stabilize and helps define the monitoring period needed to infer biofilm hydraulic properties from hydrogen-concentration measurements. This timescale is important for understanding whether the transient behavior is observable in the field measurements.

\begin{figure}[htbp]
\centering
\includegraphics[width=14cm]{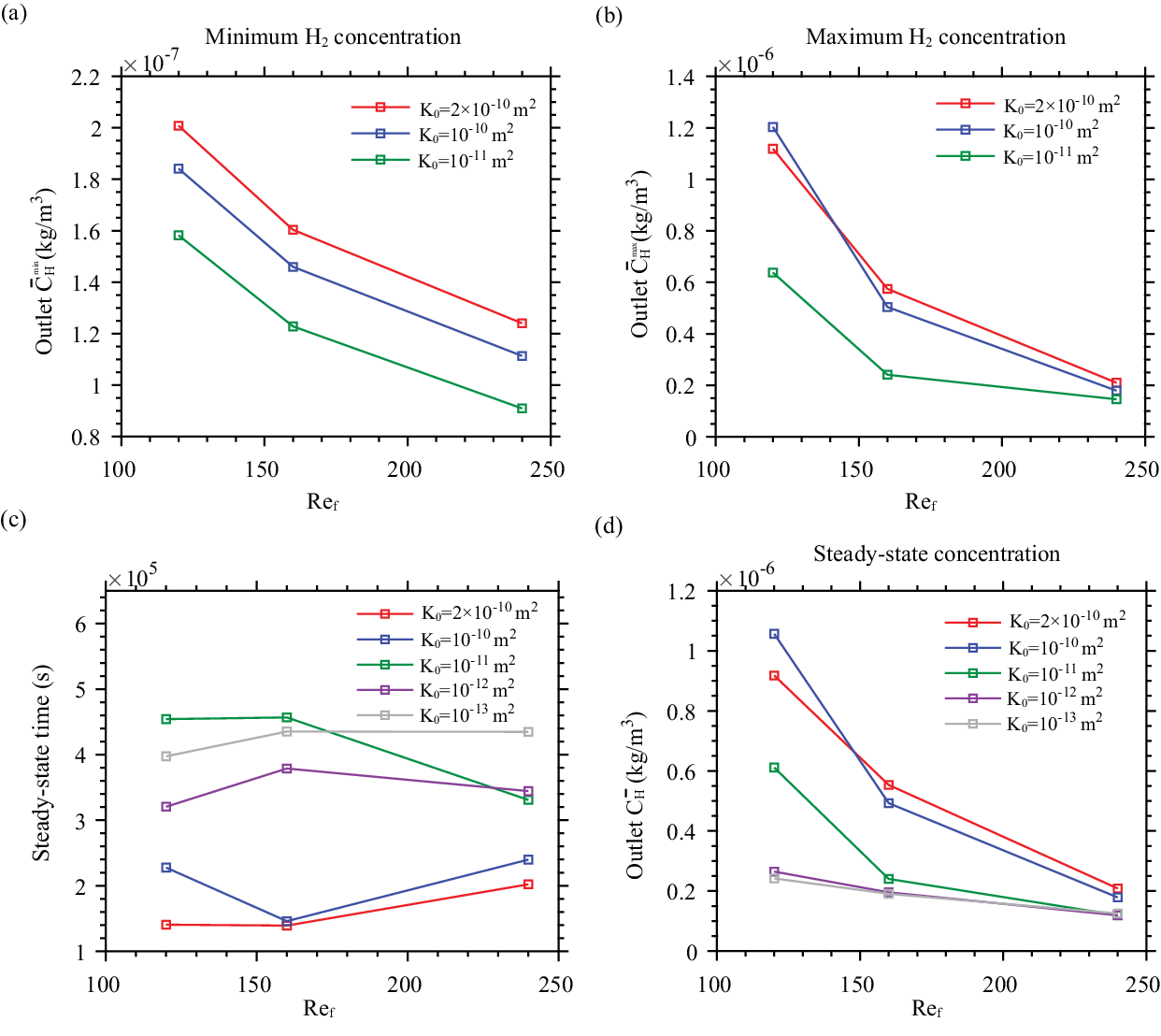}
\caption{Characteristic outlet hydrogen concentration responses. Type I: (a) Minimum concentration $\overline{C}_{H}^{min}$ (b) Maximum concentration $\overline{C}_{H}^{max}$. Type I and II: (c) Steady-state time, (d) Steady-state $\overline{C}_{H}$}
\label{fig9}
\end{figure}

For Type I, the hydrogen concentration evolves non-monotonically with time, exhibiting two characteristic values ($\overline{C}_{H}^{min}$ and $\overline{C}_{H}^{max}$). As shown in Fig.~\ref{fig9}, $\overline{C}_{H}^{min}$ decreases with increasing $Re_f$, primarily due to stronger local vortex trapping at higher $Re_f$, which significantly hinders hydrogen transport. $\overline{C}_{H}^{min}$ increases with increasing biofilm permeability $K_0$, as higher permeability facilitates hydrogen transport through the biofilm. Regarding the characteristic timescales, as discussed in Fig.~\ref{fig6}, the time required to reach $\overline{C}_{H}^{min}$ is governed by the emergence of vortex structure, and the time such a structure can exist. This structure reduces advective transport and traps the hydrogen in the reactive wall. As a result, $\overline{C}_{H}^{min}$ decreases with $Re_f$ because higher $Re_f$ can sustain the vortex structure in the fracture. Similarly, $\overline{C}_{H}^{max}$ decreases with increasing $Re_f$, as higher flow velocities reduce the fluid residence time within the reactive fracture.

For Type II, the outlet hydrogen concentration exhibits an approximately decreasing trend and reaches a quasi-steady state after a characteristic stabilization timescale. As shown in Fig.~\ref{fig9}(c), the time required to approach this state corresponds approximately to the time at which total biofilm mass reaches its maximum and the fracture flow geometry becomes nearly stationary (Fig.~\ref{fig4}). This stabilization timescale increases with $Re_f$, despite the smaller quasi-steady biofilm thickness at higher flow rates. This behavior may result from the competing effects of enhanced reactant transport and stronger hydrodynamic shear, which promote microbial growth and biofilm removal. This stabilization timescale is generally longer for Type II than for the highly permeable Type I cases. However, the timescales become comparable for $K_0=10^{-11}~\mathrm{m^2}$, which is near the transition between the two regimes.  This timescale indicates that the monitoring window for the transient behavior of Type I is about $10^5~s$ (approximately 1 day). Beyond this stage, the hydrogen concentration does not remain strictly constant but instead increases gradually as the fraction of active biofilm $\eta$ slowly decreases, thereby reducing microbial hydrogen consumption. The steady-state hydrogen concentrations for both Type I and Type II are also reported in Fig.~\ref{fig9}(d) to characterize their dependence on flow conditions and biofilm permeability. At low $Re_f$, the difference in the steady-state $\overline{C}_{H}$ between the highest $K_0$ and lowest $K_0$ is approximately four times. This difference decreases with increasing $Re_f$, consistent with the increasing dominance of the strong hydrodynamic shear.

\section{Conclusion}
\label{sec:conclusion}

In this study, we investigated the coupled processes of fracture flow, biofilm dynamics, and hydrogen transport in a rough, hydrogen-generating fracture using high-fidelity numerical simulations. The model accounts for both suspended and attached microbial biomass, including biofilm growth, shear-induced detachment and spreading. Fluid flow in the open fracture and porous biofilm is modeled using the Darcy-Brinkman-Stokes equations within a single computational domain, with local porosity and permeability updated as biofilm grows. Hydrogen is generated at the reactive fracture wall at rates representative of stimulated geologic hydrogen production.

Our results demonstrate that the role of biofilm growth goes beyond hydrogen consumption and reduction in effective fracture aperture. Biofilm growth also redistributes the flow field and limits transport between the hydrogen-generating surface and bulk fluid. Our simulation results show that biofilm growth can substantially alter flow and mass transport in the fracture, reducing the apparent fracture permeability by up to approximately $65\%$. Higher fracture Reynolds number ($Re_f$) limits biofilm growth due to shear-induced detachment, resulting in a less steady-state biofilm accumulation and smaller permeability reduction. Additionally, higher biofilm permeability ($K_0$) enhances the transport of hydrogen and other substrates through the biofilm matrix and contributes to more rapid biofilm development.

We identified two distinct regimes of hydrogen concentration evolution that are controlled primarily by intrinsic biofilm permeability. For highly permeable biofilms (Type I), the hydrogen concentration evolves non-monotonically as roughness-induced vortices initially restrict hydrogen transport and are subsequently suppressed by biofilm accumulation, leading to a transient increase in hydrogen concentration. For low-permeability biofilms (Type II), hydrogen transport is dominated by diffusion through the biofilm and microbial consumption, resulting in an approximately monotonic decrease in outlet hydrogen concentration. The transition between these regimes can be characterized by the Darcy-scale Reynolds number ($Re_D=u\sqrt{K_0}/\nu$) with a threshold of approximately $Re_D=0.5$ for the conditions investigated. Although the precise threshold may depend on fracture geometry and model parameters considered, the distinct temporal signatures of hydrogen concentration suggest that outlet hydrogen concentration measurements, with information on fracture flow conditions, may help quantify intrinsic biofilm permeability. For Type I, the characteristic transient response occurs over approximately $10^5$ s (about one day), providing a timescale for concentration monitoring. The subsequent steady-state outlet concentration may also help constrain intrinsic biofilm permeability, particularly at lower $Re_f$, where the differences among permeability cases are more significant. These temporal and steady-state responses may provide useful information on intrinsic biofilm permeability, which remains difficult to measure directly under subsurface conditions.

Finally, the current scope of this study is limited to a single fracture and constant flow rates. Future work should examine a fracture network with various fracture geometries and time-varying flow rates. Incorporating these coupled processes into reservoir-scale models will be important for assessing how microbial activity influences hydrogen production in stimulated geologic hydrogen systems.

\section*{\textbf{Declaration of Competing Interest}}
The authors declare that they have no competing interests. 

\section*{\textbf{Declaration of generative AI and AI-assisted technologies in the manuscript preparation process}}
During the preparation of this work, the authors used ChatGPT in order to polish sentence structures. After using this tool, the authors reviewed and edited the content as needed and take full responsibility for the content of the published article.

\section*{\textbf{Acknowledgements}}

This research was supported by the Targeted Proposal Teams (TPT) Collaborative Seed Grants program through the Division of Research at Texas A\&M University. Portions of this research were conducted with the advanced computing resources provided by Texas A\&M High Performance Research Computing.

\bibliographystyle{elsarticle-num} 

\end{document}